\documentclass[12pt,dvips]{article}
\usepackage{graphicx}
\usepackage{color}
\usepackage{txfonts}
\usepackage{natbib}
\begin{document}
\title{Atmospheric mass loss by stellar wind from planets around main sequence M stars}
\author{J. Zendejas$^1$, A. Segura$^{2,*}$, A. C. Raga$^2$}
\date{}

\maketitle
\noindent  $^1$Instituto de Astronom\'{i}a, Universidad  Nacional Aut\'{o}noma de M\'{e}xico. A.P. 70-264, 04510 D.F., M\'{e}xico; $^2$Instituto de Ciencias Nucleares, Universidad  Nacional Aut\'{o}noma de M\'{e}xico. A.P. 70-543, 04510 D. F., M\'exico; $^*$member of the Virtual Planetary Laboratory a project of the NASA Astrobiology Institute.

\vspace{12pt}
\noindent Pages: 29

\noindent Figures: 6

\noindent Tables: 0

\vspace{12pt}           
\noindent\textbf{Proposed Running Head:} Atmospheric mass loss on planets around M dwarfs

\vspace{12pt}
\noindent\textbf{Editorial correspondence to:}

Jes\'us Zendejas

Present address:

Max Planck Institute for Astrophysics, Karl-Schwarzschild-Str. 1

85741 Garching, Germany

Ph: +49-89-30000-0

Fax:+49-89-30000-2235

E-mail address: chicho@usm.lmu.de, jesusz@astroscu.unam.mx

\newpage
\begin{abstract}
   
  {We present an analytic model for the interaction between planetary
atmospheres and stellar winds from main sequence M stars, with the purpose of
obtaining a quick test-model that estimates the timescale for total
atmospheric mass loss due to this interaction. Planets in the habitable zone of M dwarfs may be tidally locked and may have weak magnetic fields, because of this we consider the extreme case of planets with no magnetic field.
The model gives the planetary atmosphere mass loss rate as a function of the stellar wind and planetary properties (mass, atmospheric pressure and orbital distance) and an entrainment efficiency coefficient $\alpha$. 
We use a mixing layer model to explore two different cases: a time-independent stellar mass loss and a stellar mass loss rate that decreases with time. For both cases we consider  planetary masses within the range of $1\to10$ M$_{\oplus}$ and
atmospheric pressures with values  of 1, 5 and 10 atm. We apply our model to Venus by estimating its atmospheric mass loss rate by the interaction with the solar wind and compare our model with more detailed simulations. We find a good agreement between our results and the atmospheric mass loss obtained by more detailed models, and it is therefore appropriate for carrying out an exploration of the broad parameter space of exoplanetary systems. 
For the time dependent case, planets without magnetic field in the habitable zone of M dwarfs with initial stellar mass losses of
  $\leq \dot{M}_{w} < 10^{-11}$ M$_{\odot}$ yr$^{-1}$, may retain their atmospheres for at least 1 Gyr. This case may be applied to early spectral type M dwarfs (earlier than M5). Studies have shown that late type M dwarfs (later than M5) may be active for long periods of time ($\geq 4$Gyr), and because of that our model with constant stellar mass loss rate may be more accurate. For these stars most planets may have lost their atmospheres in 1 Gyr or less because most of the late type M dwarfs are expected to be active. We emphasize that our model only considers planets without magnetic fields. Clearly we must expect a higher resistance to atmospheric erosion if we include the presence of a magnetic field. Nevertheless, as a first approximation our model is able to give a reliable time scale, as evidence by comparing our results with more detailed models.}

\end{abstract}

\newpage

\section{Introduction}

Low-mass main sequence stars of spectral class M (M dwarfs, or dM stars) are the most abundant stars in the galaxy,
representing about 75\% of the total stellar population. Their low masses                       
(0.08 - 0.8 M$_{\odot}$) allow them to have main sequence lifetimes of the order of $10^{11}$ years (or more), i.e.,
significantly longer than the age of the universe. During this time the dM stars present 
nearly constant luminosities ($L$ $\sim10^{-1}-10^{-3}$L$_{\odot}$, \citealt{Scalo1}).
These peculiar characteristics make them interesting systems to search for habitable planets. However, there are some issues that
make the habitability of planets around these stars a matter of
debate. We will not review here the habitability issues related to
planets around M stars,  referring the reader to the papers by
\cite{Tarter} and  \cite{Scalo1}.

M stars exhibit significant temporal variability as a consequence of phenomena
occurring in the region from their photospheres to their coronae.
As a result, they emit large amounts of ultraviolet (UV) radiation and X-rays during their
active periods and strong stellar winds. This stellar activity can have several effects on a planet. In this paper we
focus on the interaction of stellar winds with the atmosphere of
planets around these stars.
Stellar winds are mainly charged particles (protons and
electrons) ejected from the outer stellar atmosphere driven by the
pressure-expansion of the hot corona. 
It has been calculated that the M dwarf stellar winds
are denser and faster than winds from solar-like stars \citep{Wargelin}, and
these winds may be able to erode the atmospheres of planets around
the stars.

The partial or total erosion of a planetary atmosphere affects directly the habitability
of a planet, since a main requirement for the development of life on its surface is the
presence of an atmosphere. The planetary atmosphere stabilizes the surface temperature and
provides the pressure for water to be liquid, an essential requirement
for life. Therefore, if a planetary atmosphere is removed by the stellar wind from its
parent star (or any other process for that matter), this represents a serious problem
for the habitability of the planet. For habitable planets around M dwarfs, the stellar wind may have a
larger effect, not only because it is
denser and faster than the winds from solar-type stars, but because their
habitable zones (HZ) correspond to orbits very near to the star.

The concept of HZ is commonly used in the context of planets
that might be suitable to keep life on their surface. There are many
definitions for the HZ ($e$.$g$ \citealt{Dole} and \citealt{Heath}) but we take
here Kasting's concept which is defined as the region around a star in which a planet with an atmosphere is able to
support liquid water on the surface  \citep{Kasting}. The formal definition is related to the stellar luminosity by:
\begin{equation}
\centering
0.95AU\leqslant\frac{D}{\sqrt{{L_*}/{L_\odot}}} \leqslant1.37AU
\end{equation}
where $D$ is the distance from the planet to the star, $L_*$ is the stellar luminosity and L$_\odot$ is the solar luminosity. For the M dwarf case this distance is $\sim$0.2 AU from the star. Planets in the HZ of M dwarfs would become tidally locked in about 1 Gyr or less \citep{Griessmeieretal}. Calculations indicate that the magnetic moment of a tidally locked terrestrial planet may be 10 to 100 less intense than Earth's magnetic moment. Because it is likely that planets in the HZ of M dwarfs have weak magnetic fields, here we assume the extreme case where the planet has no intrinsic magnetic field, like Venus.

The search for extrasolar planets is an interesting issue in the context of Astrobiology.
The smaller planets have been found orbiting dM stars, due to the fact that it is easier to observe exoplanets with the current techniques (gravitational lensing, radial velocities or transits) around less luminous stars. Since it is potentially possible to detect habitable planets around M dwarf stars, the study of the habitability conditions of these planets is decisive in the search for life around dM stars. Some recent exoplanet searches such as CoRoT (COnvection, ROtation \& planetary Transits) \citep{Aigrain2008,Pierre2008} and HARPS (High Accuracy Radial velocity Planetary Search project) \citep{Udry2007, Pepe2000} are dedicated to looking for planets with Earth-like masses around M dwarfs.             
\section{Atmospheric mass loss calculation}
\subsection{Erosion of the atmosphere by the stellar
wind}\label{constantrate}

We first carry out an estimate of the atmospheric mass loss due to
the interaction of a stellar wind of density $\rho_w$ and velocity
$v_w$ with a planetary atmosphere. For an Earth-type planet, the atmosphere
is a thin, gaseous layer covering the planet. Therefore, both the inner
and outer radii of the atmosphere are $\approx R_P$, where $R_P$ is
the radius of the planet.

For a planet with no magnetic field, the impinging stellar wind will
form a bow shock around the body of the planet and its atmosphere. 
For low enough stellar wind densities this bow shock will not exist and the stellar wind will be directly absorbed by the planetary atmosphere. For a stellar wind/planet interaction in which the bow show does exist the
post-bow shock material will have a sound speed $c_s\sim v_w$ (as the
shock will be strong due to the hypersonic nature of the stellar wind).
The post-bow shock flow will envelop the leading hemisphere
of the planet, and a turbulent mixing layer will be formed on the contact
surface between the shocked wind and the planetary ionosphere. The
shocked wind is stationary in the stagnation region of the bow shock,
and has an increasing velocity towards the bow shock wings, reaching
velocities of $\sim v_w$  towards the edge of the leading hemisphere
of the planet/wind interaction flow. 

From the mixing layer formalism of \citet{CantoRaga} it is
straightforward to show that a flow with a sound speed $\sim v_w$ and
a tangential velocity of the same order has an entrained mass flux is
${\dot M} \approx \alpha \rho_w v_w$ (where $\alpha$ is an entrainment
efficiency which is
determined through fitting the mixing layer model to laboratory experiments).
Multiplying the entrained mass flux by the surface of the leading
hemisphere, we then obtain an estimate for the atmospheric mass per unit
time entrained by the stellar wind:
\begin{equation}\label{loss}
\dot{M}_a \approx 2 \pi R_P^2 \alpha \rho_w v_w
\end{equation}
where $\dot{M}_a$ represents the atmospheric mass loss. This equation
can be combined with the relation ${\dot M}_w=4\pi D^2\rho_wv_w$ (where
${\dot M}_w$ is the stellar mass loss rate and $D$ is the orbital radius of the planet) to obtain:
\begin{equation}
\dot{M}_a=\left({R_P\over D}\right)^2{{\dot M}_w\alpha\over 2}\,.
\label{tloss}
\end{equation}

Through comparisons with laboratory experiments of plane, turbulent mixing
layers, \citet{CantoRaga} determined an $\alpha=0.03$ entrainment
efficiency. It
is of course not clear whether or not this value for $\alpha$ is appropriate
for the different flow geometry of our stellar wind/planetary atmosphere
interaction model. Nevertheless, \citet{Bauer2004} suggested that the value of $\alpha \approx 1/3$ for the case of Venus. In order to sample the possible range of $\alpha$ values we use $\alpha=$0.01 and 0.3.

We should note that the interaction between the solar wind and Venus is found to be more complex than we have described above. For example, a magnetic barrier (of compressed, magnetized solar wind material) is formed in the solar wind/planetary ionosphere interface \citep{Russelletal}. This phenomenon could modify the properties of the entrainment process in a substantial way. Therefore, our entrainment model should be considered as a parametrization with which a range of possible entrainment rates can be explored regardless of the physical constraints for the $\alpha$ parameter.

\subsection{The timescale for total atmospheric erosion}\label{variable}
In order to estimate the initial atmospheric mass $M_{atm}$ as
a function of the surface atmospheric pressure $P_0$, we use
the hydrostatic equation for an isothermal atmosphere in the form:
\begin{equation}\label{equation6}
M_{atm}=\frac{4 \pi P_0 R_P^2}g\,,
\end{equation}
where $P_0$ is the surface pressure, $R_P$ the planetary radius
and $g$ the gravitational acceleration at the planetary surface.
This equation can be
combined with the relations $g=GM_P/R_P^2$ (where $G$ is the gravitational
constant and $M_P$ the planetary mass) and $M_P=4\pi R_P^3\rho_P/3$
(where $\rho_P$ is the average density of the planet) to obtain
\begin{equation}
M_{atm}=\left({3\over \rho_P}\right)^{4/3}{P_0 M_P^{1/3}\over
(4\pi)^{1/3}G}\,.
\label{matm0}
\end{equation}
We now consider planets with different masses $M_P$,
assuming the Earth average density (i.e.,
$\rho_P=5.5$g~cm$^{-3}$), and from Eq. (\ref{matm0}) we
obtain~:
\begin{equation}
{M_{atm}\over M_{a,\oplus}}=\left({P_0\over {1\,{\rm atm}}}\right)
\left({M_P\over M_\oplus}\right)^{1/3}\,,
\label{matm}
\end{equation}
where 1~atm=$10^6$g~cm$^{-1}$~s$^{-2}$, $M_\oplus=5.97\times 10^{27}$~g
and $M_{a,\oplus}=5.27\times 10^{21}$~g are the Earth's surface
atmospheric pressure, planetary mass and atmospheric mass, respectively.

From Eqs. (\ref{tloss}) and (\ref{matm}) we can then obtain the characteristic
timescale $t_0$ for total loss of the planetary atmosphere~:
\begin{equation}
t_0\equiv {M_{a,0}\over {\dot M}_a}=
\left({P_0\over {\rm 1\,atm}}\right)\left({M_\oplus\over M_P}\right)^{1/3}
\left({D\over {\rm 0.2\,AU}}\right)^2\left({0.3\over \alpha}\right)
\left({{\dot M}_\odot\over {\dot M}_w}\right)\times 4\times 10^{10}\,
{\rm yr}\,,
\label{t0}
\end{equation}
normalized to $D = 0.2$~AU, the average  distance for a planet in the
habitable zone around an M dwarf.

If the mass loss rate of the star ${\dot M}_w$ is time-independent,
the timescale $t_0$ is equal to the time for total erosion of the
planetary atmosphere. From Eq. (\ref{t0}) we then conclude that
the timescale for total atmospheric  mass loss will satisfy the
condition  $t_0\geq 1$~Gyr provided that
\begin{equation}
{{\dot M}_w\over {\dot M}_\odot}\leq 40
\left({P_0\over {\rm 1\,atm}}\right)\left({M_\oplus\over M_P}\right)^{1/3}
\left({D\over {\rm 0.2\,AU}}\right)^2\left({0.3\over \alpha}\right)\,.
\label{cond0}
\end{equation}
Therefore,
for an atmosphere with initial surface pressure $P_0=1$~atm of
a planet with $M_P=1\,M_\oplus$ and an orbital radius
$D=0.2$~AU, the timescale for total atmospheric loss will be long
enough ($\approx 1$~Gyr) for the development of life if
${\dot M}_w\leq 40\,{\dot M}_\odot$, if we assume an
$\alpha=0.3$ entrainment efficiency. Here ${\dot M}_\odot$ is the present solar mass loss, 10$^{-14}$ M$_\odot$yr$^{-1}$. For a lower, $\alpha=0.01$
entrainment efficiency (see \S 2.1), we derive a higher,
${\dot M}_w\leq 1200 \,{\dot M}_\odot$ value for the upper limit
of the allowed stellar mass loss rate.

\subsection{Atmospheric loss due to a time-dependent stellar wind}

Observations indicate that the stellar mass loss rate has a temporal
dependence for solar-type stars 
\citep{Wood}. We now assume a similar dependence on the loss
rate for main sequence M stars. \citet{Wood} obtained a
temporal dependence following the power law
\begin{equation}
\dot{M}_{w} = {\dot{M}_{w,0}} \left({{t_w}\over{t+t_w}}\right)^{2}\,,
\label{power}
\end{equation}
where  $t_w=0.1$~Gyr and $\dot{M}_{w,0}$ is the stellar mass loss rate
at $t = 0$. For the Sun, ${\dot{M}_{\odot,0}}=2\times10^{-11}$
M$_{\odot}$ yr$^{-1}$ \citep{Wood}. In other words,
in its early main sequence lifetime, the Sun had a mass loss rate
$\sim 2000$ times larger than its present mass loss rate (10$^{-14}$ M$_\odot$yr$^{-1}$).

We now combine Eqs. (\ref{tloss}) and (\ref{power})
to obtain:
\begin{equation}
{\dot M}_a=-{dM_a\over dt}=\left({R_P\over D}\right)^2
\left({\alpha {\dot M}_{w,0}\over 2}\right){t_w^2\over {(t+t_w)^2}}\,.
\label{int}
\end{equation}
This equation can be integrated over time. Using
Eq. (\ref{tloss}) and the definition of Eq. (\ref{t0})
($t_0 \equiv M_{a,0}/\dot{M}_a$), we obtain:
\begin{equation}
M_a(t)=M_{a,0}\left[1-\left({t_w\over t_0}\right)
{t/t_w\over t/t_w+1}\right]\,,
\label{mat}
\end{equation}
where $M_{a,0}$ is the initial mass of the planetary atmosphere
and $t_0$ is given by Eq. (\ref{t0}) with ${\dot M}_w={\dot M}_{w,0}$.
By setting $M_a(t_f)=0$ in Eq. (\ref{mat}) we then obtain the
time
\begin{equation}
t_f=t_w {t_0/t_w\over {1-t_0/t_w}}\,,
\label{tf}
\end{equation}
for total atmospheric mass loss. In Fig. 1 we plot $t_f$ as a function
of $t_0/t_w$. We have $t_f\approx t_0$ for $t_0\ll t_w$ (= 0.1 Gyr,
see above) and $t_f\to \infty$ for $t_0\to t_w$.

From Fig. 1, we see that Eq. (\ref{tf}) gives $t_f/t_w\approx 10$
for $t_0/t_w\approx 0.9$ (in other words, for $t_0=
0.9\times 0.1$~Gyr). Therefore, $t_f\geq 1$~Gyr (the timescale
necessary for the development of life) is obtained for
$t_0\geq 0.09$~Gyr. Using this limit in Eq. (\ref{t0}),
we obtain the condition
\begin{equation}
{{\dot M}_{w,0}\over {\dot M}_\odot}\leq 433
\left({P_0\over {\rm 1\,atm}}\right)\left({M_\oplus\over M_P}\right)^{1/3}
\left({D\over {\rm 0.2\,AU}}\right)^2\left({0.3\over \alpha}\right)\,.
\label{cond1}
\end{equation}
This condition allows for stellar wind
mass loss rates a factor of $\sim 10$ higher
than Eq. (\ref{cond0}), which was derived assuming a time-independent
stellar mass loss rate.

\begin{figure}
   \centering
   \includegraphics[angle=90,width=12cm]{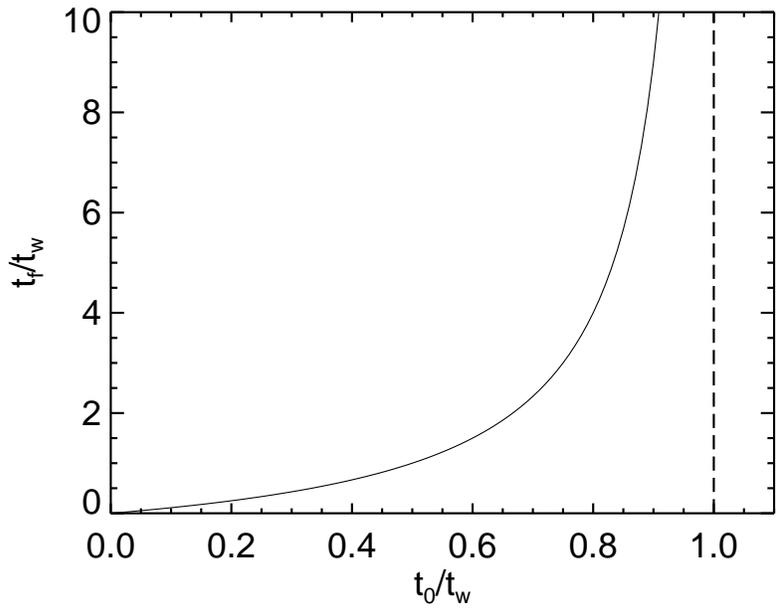}
      \caption{Total atmospheric erosion time $t_f$ resulting from
the evolving stellar wind model (see Eq. \ref{tf}) as a function
of $t_0$ (which is the total atmospheric loss time obtained assuming
a time-independent stellar wind, see Eq. \ref{t0}). Both $t_f$ and
$t_0$ are given in units of $t_w=0.1$~Gyr (see Eq. \ref{power}).
              }
         \label{tlossfig}
   \end{figure}

\section{Conditions for atmospheric survival obtained
from the time-independent and time-dependent
stellar wind models}

Figures 2, 3, 4, and 5 show the stellar mass loss necessary
for a planet without magnetic field located at 0.2 AU to conserve its atmosphere for 1 Gyr,
considering the time independent and time-dependent stellar wind models.
Lower stellar mass loss rates than the ones given in these figures result in longer times for total loss of the planetary atmospheres
(Eqs. \ref{cond0} and \ref{cond1}). The mass loss rates shown in
these figures have been computed from Eqs. (\ref{cond0}) and (\ref{cond1})
assuming  the $\alpha= 0.01$ entrainment coefficient (in the bottom of the range implied by laboratory
experiments, see \S 2.1), the obtained mass loss rates are plotted in the  Figs. 2 and 4. The results for $\alpha=0.3$ entrainment coefficient are presented in Figs. 3 and 5.

As expected, planets with
larger initial atmospheric pressures have more resistance to erosion.
More massive planets loose their atmospheres faster than smaller
planets. This result is due to the fact that (for a set of planets with the same density $\rho_p$, 
see Eq. \ref{variable}) while the surface area presented to the stellar wind increases as $M_P^{2/3}$, the total
atmospheric mass increases only as $M_P^{1/3}$ (see Eq. \ref{matm0}).

From Figs. 2 to 5 we see that from the
time-dependent stellar wind model (three upper lines) stellar mass
loss rates required for totally loss a planetary atmosphere within 1 Gyr, are an order of magnitude higher
than the ones obtained from the time-independent stellar wind model (three
lower lines). These results imply that the atmosphere of a planet in the habitable zone
of an active M dwarf may have an atmosphere that lasts for $\approx 1$Gyr
only if the two following conditions are met~:
\begin{enumerate}
\item the entrainment parameter $\alpha$ is close to $\alpha=0.01$
\item a substantial drop in the stellar wind mass loss rate occurs
over a timescale of $\sim 1$~Gyr (in other words, that the mass loss
rate falls at a rate comparable to the one assumed in our time-dependent
stellar wind model, see \S 2.3).
\end{enumerate}

If these conditions are not met, only a planetary atmosphere with a very
high initial atmospheric pressure may survive for timescales of $\sim
1$~Gyr under the action of a stellar wind with
$\dot{M}_{w} \approx 10^{-10}$ M$_{\odot}$ yr$^{-1}$.

\begin{figure}
   \centering
   \includegraphics[angle=90,width=12cm]{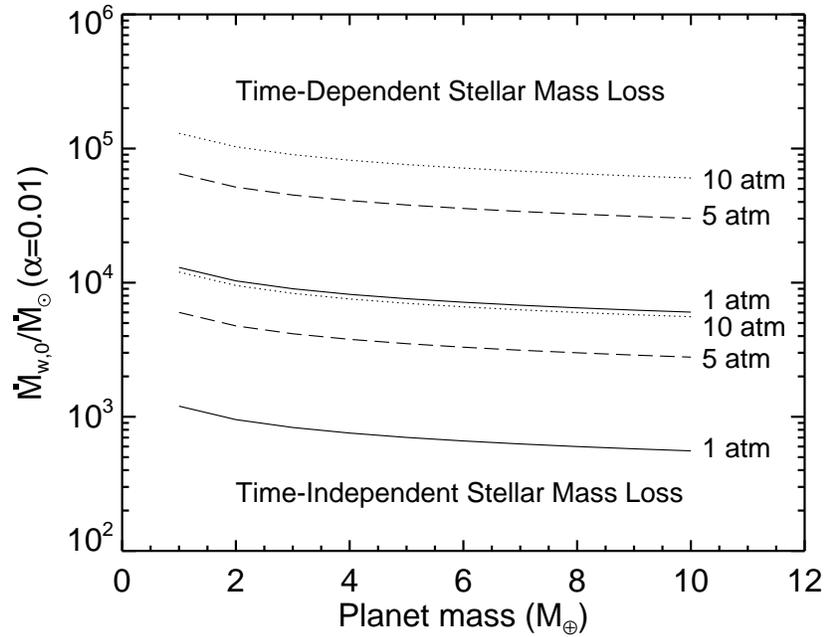}
\caption{Initial stellar mass loss rate needed in order to keep a planetary
atmosphere for 1 Gyr in the habitable zone (at 0.2 UA) of a M dwarf,
for planets with different masses. The three upper lines
have been computed with the time-dependent wind model, and
the three lower lines with the time-independent wind model.
This figure corresponds to $\alpha=0.01$ and the solar mass loss 
considered here is for the current Sun value
($\dot{M}_{\odot} \approx 10^{-14}$ M$_{\odot}$ y$r^{-1}$).}
\label{figmass_01}
\end{figure}

\begin{figure}
   \centering
   \includegraphics[angle=90,width=12cm]{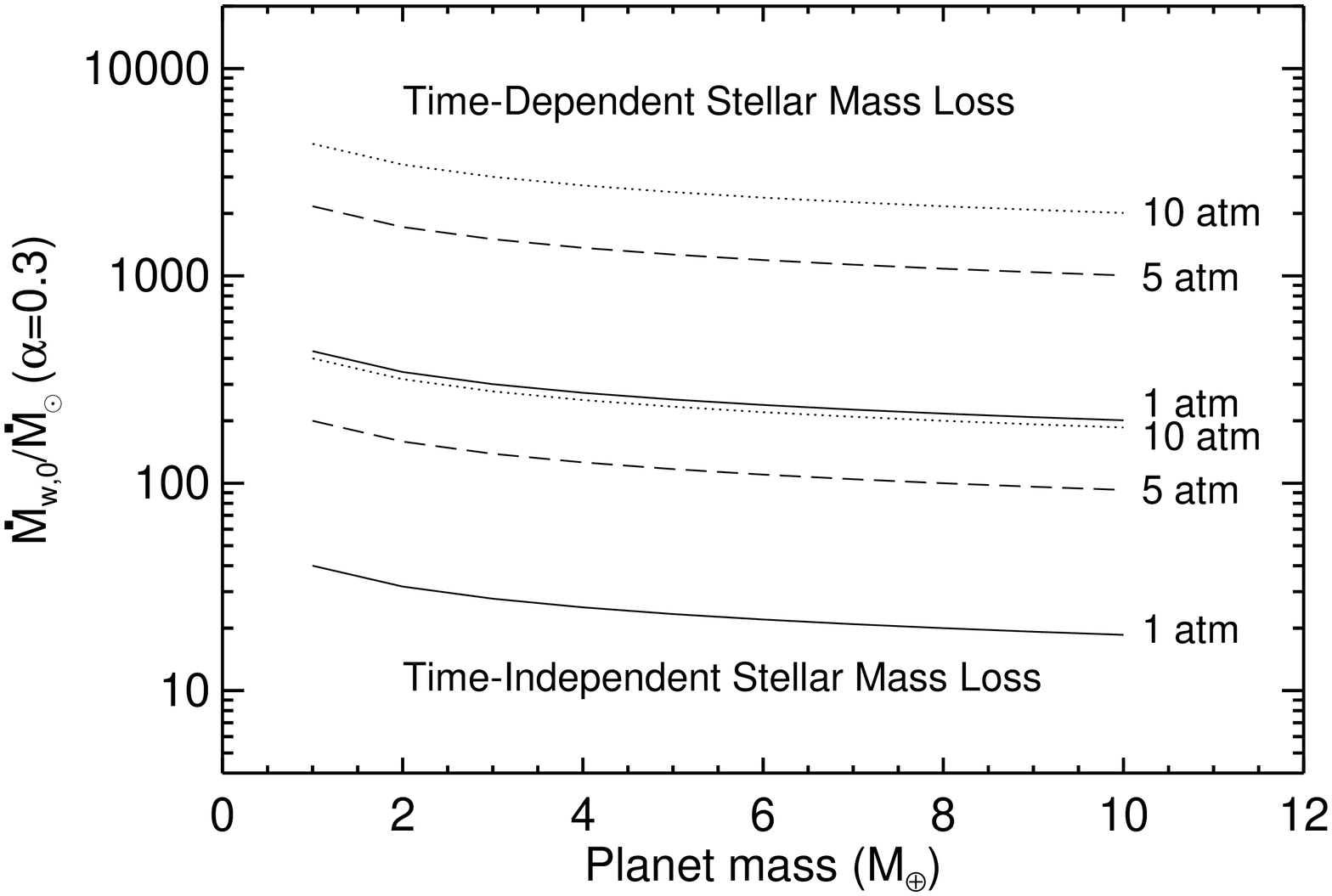}
\caption{Initial stellar mass loss rate needed in order to keep a planetary
atmosphere for 1 Gyr in the habitable zone (at 0.2 UA) of a M dwarf,
for planets with different masses. The three upper lines
have been computed with the time-dependent wind model, and
the three lower lines with the time-independent wind model.
This figure corresponds to $\alpha=0.3$ and the solar mass loss 
considered here is for the current Sun value
($\dot{M}_{\odot} \approx 10^{-14}$ M$_{\odot}$ y$r^{-1}$).}
\label{figmass_03}
\end{figure}

\begin{figure}
\centering
\includegraphics[angle=90,width=12cm]{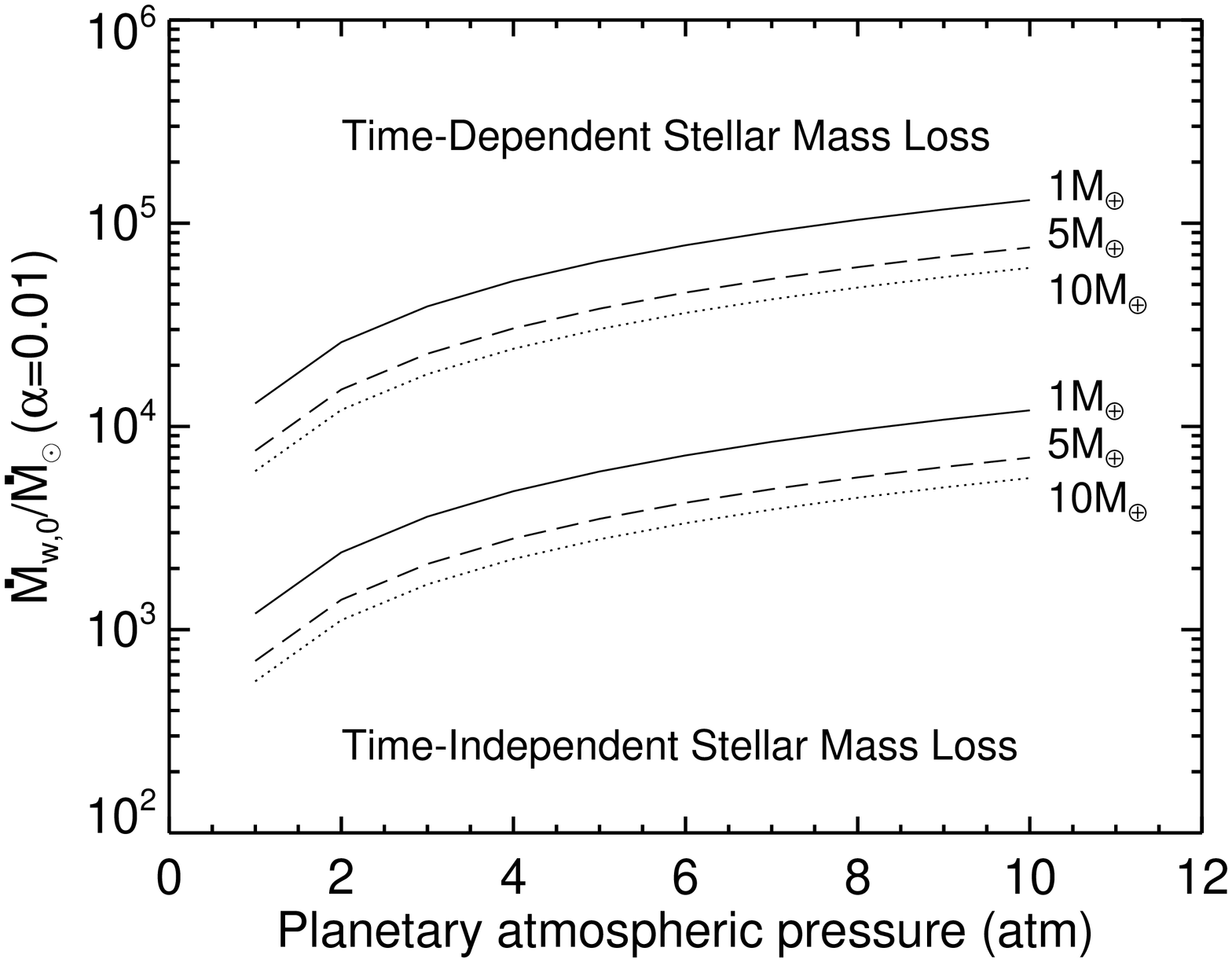}
\caption{Initial stellar mass loss rate needed in order to keep a planetary
atmosphere for 1 Gyr in the habitable zone (at 0.2 UA) of a M dwarf,
for planets with different atmospheric pressures. The three upper lines
have been computed with the time-dependent stellar wind model, and
the three lower lines with the time-independent wind model.
This figure corresponds to $\alpha=0.01$ and the solar mass loss 
considered here is for the current Sun value
($\dot{M}_{\odot} \approx 10^{-14}$ M$_{\odot}$ y$r^{-1}$).}
\label{figpress_01}
\end{figure}

\begin{figure}
\centering
   \includegraphics[angle=90,width=12cm]{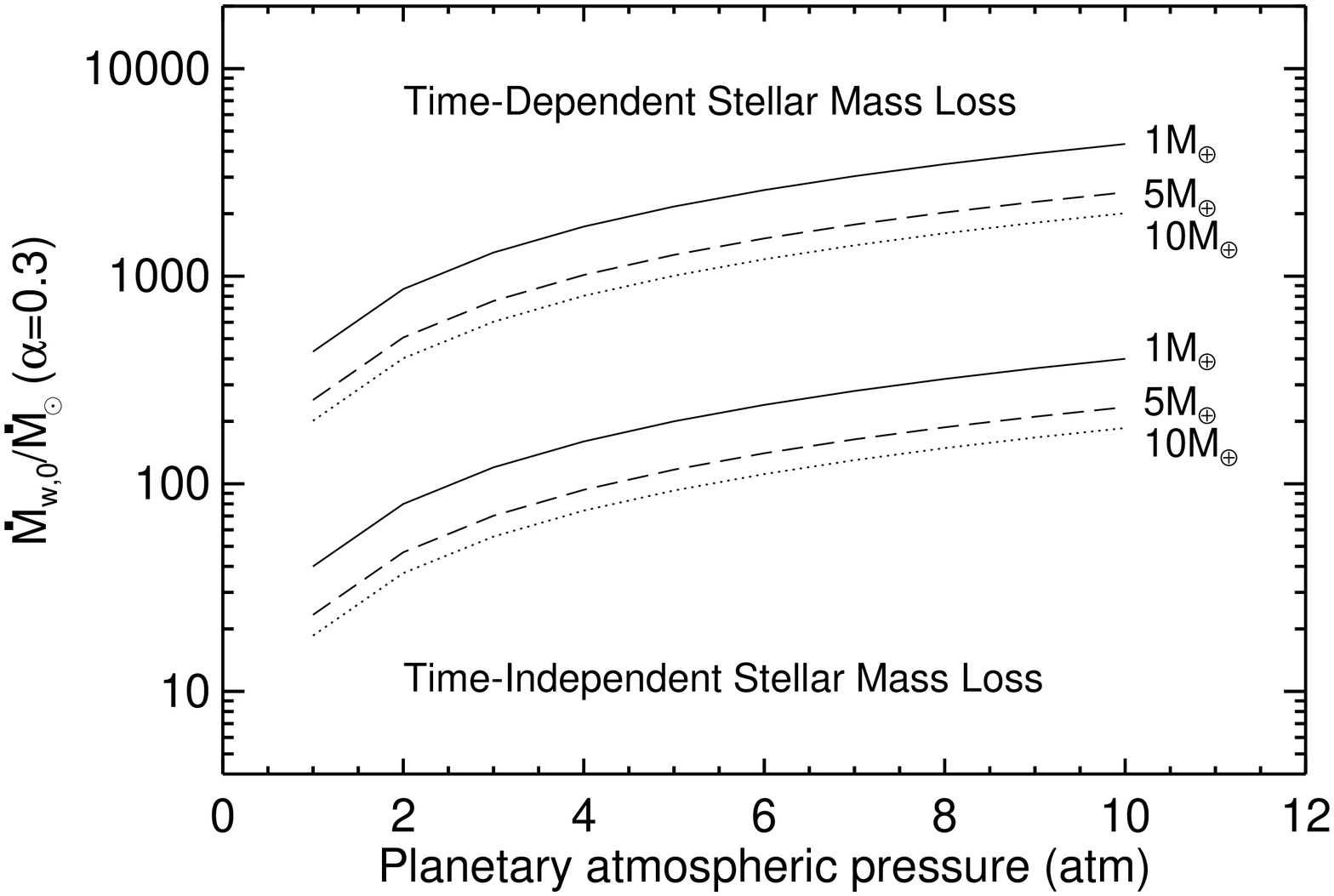}
\caption{Stellar mass loss rate needed in order to keep a planetary
atmosphere for 1 Gyr in the habitable zone (at 0.2 UA) of a M dwarf,
for planets with different atmospheric pressures. The three upper lines
have been computed with the time-dependent stellar wind model, and
the three lower lines with the time-independent wind model.
This figure corresponds to $\alpha=0.3$ and the solar mass loss 
considered here is for the current Sun value
($\dot{M}_{\odot} \approx 10^{-14}$ M$_{\odot}$ y$r^{-1}$).}
\label{figpress_03}
\end{figure}

\subsection{Venus comparison}
Several authors have suggested that Venus may have lost
a considerable amount of its atmosphere trough its interaction with the solar wind. We used our time-dependent stellar wind model to calculate the atmospheric mass loss rate on the early Venus atmosphere.
  The initial solar mass loss rate was considered to start with $\sim 2\times10^{-11}$  M$_{\odot}$ yr$^{-1}$ at 0.1 Gyr \citep{Wood} and evolve with time as presented in Eq. \ref{power}. We calculated the values for the atmospheric
mass loss rate using Eq. \ref{int}.

Our results are comparable to those obtained by \citet{Kulikov}, who obtained similar values for the atmospheric mass loss rate using a numerical test particle model (Fig. 6). \citet{Kulikov} obtained series of values for the O$^+$ pickup loss, considering the interaction of stellar winds and XUV flux of the young Sun, assuming different atmospheric density profiles. They used different parameters for the solar wind integrated over time for the period from 3.5 to 4.6 Gyr before present. They suggest that Venus may have lost during 4.5 Gyr more than 250 bars of O$^+$ ions. In Fig. 6 we compare our model with cases 1b and 3a from \citet{Kulikov}. Case 1b is for the maximum solar wind and for the obstacle boundary that is one oxygen scale height below the exobase level, and case 3a is for the minimum solar wind and for the exobase level chosen as the boundary for the solar wind-exosphere interaction. These cases were chosen because they have the maximum and minimum atmospheric mass loss rates for Venus.

Our results were also compared with other works that estimate the pickup rate for O$^+$ ions considering the present solar wind conditions. For example, \cite{Lammer} and \cite{Lammer2} use a gas dynamic test particle model to obtain an average loss rate of 1.6$\times10^{25}$ s$^{-1}$ and 1.5$\times10^{25}$ s$^{-1}$ respectively. In an earlier work, \cite{Moore} estimated a value of about $1\times10^{24}$ s$^{-1}$ applying a three-dimensional global hybrid simulation of the solar wind interaction with the dayside of Venus. Measurements carried out by the Pioneer Venus Orbiter \citep{MihalovBarnes} and Venus Mars Express \citep{Barabashetal} indicate atmospheric mass losses of $\sim 10^{25}$ ions s$^{-1}$; therefore our model is in good agreement with more detailed models and observations of the Venus atmospheric mass loss, with the plot for $\alpha=0.3$ being closer to the results obtained by other authors.

\begin{figure}
\centering
   \includegraphics[angle=90,width=12cm]{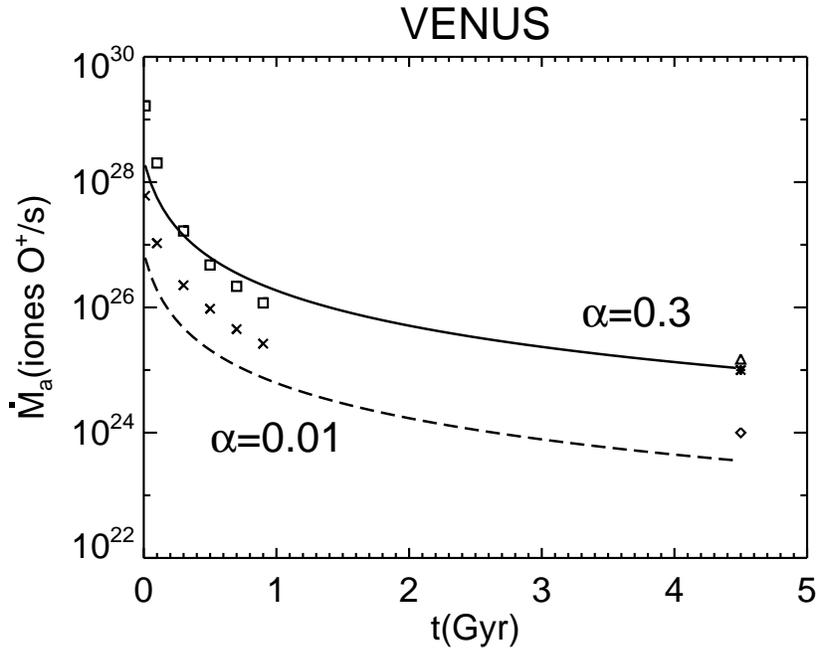}
\caption{Comparative for the atmospheric mass loss rate on the early Venus atmosphere
obtained by our time-dependent stellar wind model and more detailed models. The solid and dashed lines represent the values corresponding to our model considering values of $\alpha$=0.3 and 0.01 respectively. Squares and crosses show the values obtained by \cite{Kulikov} for their cases 1b and 3a, respectively.  Rhombus, triangle and star represent the values calculated by  \cite{Moore}, \cite{Lammer} and \cite{Lammer2}, respectively. See text for more details.}
\label{venus}
\end{figure}

\subsection{Discussion: Survival of planetary atmospheres in the
habitable zones of M dwarfs}

In general, there is little observational data for M dwarf winds. The first work on this topic
was made by \citet{Wright}, who pointed out that the stellar mass
loss may be inferred from observations in the radio region of the spectrum.
Following this idea, \citet{Mullan92} used a low frequency test with the James Clerk Maxwell Telescope ($JCMT$)
and the Infrared Astronomical Satellite ($IRAS$). Their results suggested a stellar mass loss rate of a few
times $\sim$ $10^{-10}$ M$_{\odot}$ yr$^{-1}$, four orders of magnitude larger
than the present solar mass loss. Later, \citet{LimWhite1996} used observations of the dMe stars YZ CMi and AD Leo
at the millimetric wavelength range to show that the mass loss rate of this type of stars must be less than  $\sim$ $10^{-13}$ M$_{\odot}$ yr$^{-1}$. 

\citet{VandenOortDoyle} analyzed the data of $JCMT$ and $IRAS$ reaching a different conclusion than \citet{Mullan92}. By solving the radiative transfer equations for stellar winds from dMe stars, they
showed that the inferred power-law flux distributions, based on
radio, $JCMT$ and $IRAS$ data, cannot be reconciled with the flux distributions from a stellar wind of $\sim$ $10^{-10}$ M$_{\odot}$ yr$^{-1}$. According to \citet{VandenOortDoyle} the maximum allowable mass loss rate is at most a few times 
$\sim$ $10^{-12}$ M$_{\odot}$ yr$^{-1}$, while \citet{HolzwarthJardine} derive a maximum mass loss of 10${\dot M}_\odot$, using a model for the wind properties of cool main-sequence stars, which comprises their wind ram pressures, mass fluxes, and terminal wind velocities.\citet{Wargelin} proposed a new observational technique for measuring stellar winds from late type dwarfs (F-M) through
observations of charge exchange-induced X-ray emission, but its results have been limited until now \citep{WargelinDrake2002, Wargelinetal}.

Many M dwarfs present their H$\alpha$ line in emission and because of that are classified as dMe stars. The strength of the H$\alpha$ line is considered as a proxy of activity associated to higher mass loss rates than those for stars with no H$\alpha$ emission. One of the most observed dMe stars is Proxima Centauri that has a measured mass loss rate $< 3\times 10^{-13}$ M$_{\odot}$ yr$^{-1}$ \citep{WargelinDrake2002}. Proxima Centauri is a M5.5Ve star member of a multiple system with an age of $4 - 4.5\times10^9$ years \citep{Demarqueetal}. Using this range for the age of Proxima Centauri and Eq. (\ref{power}), the initial mass loss rate of Proxima Centauri lies between 5$\times 10^{-10}$ M$_{\odot}$ yr$^{-1}$ and 6.4$\times 10^{-10}$ M$_{\odot}$ yr$^{-1}$. If Proxima Centauri is a young star ($<$ 1 Gyr) as obtained by \cite{Flemingetal}, then its initial mass loss rate would be $\sim4\times 10^{-11}$ M$_{\odot}$ yr$^{-1}$. For the case of other dMe stars that are considered young, like YZ CMi (dM4.5e) and AD Leo (dM3.5e) \citep{Flemingetal}, the calculated maximum for their mass loss rate is $10^{-12}$ M$_{\odot}$ yr$^{-1}$ \citep{VandenOortDoyle}. The age of these stars lies somewhere between 0.5 Gyr \citep{Soderblom} and 1 Gyr \citep{Youngetal}. Therefore their initial mass loss rates may be in the range of 1.1$\times 10^{-11}$ M$_{\odot}$ yr$^{-1}$ to 3.6$\times 10^{-11}$ M$_{\odot}$ yr$^{-1}$. YZ CMi and AD Leo are among the most active M dwarfs, and  we would therefore expect that they represent an upper limit for M dwarf mass loss rates.

Here we are considering that the mass loss rate for dM stars is time dependent, but observations indicate that some M dwarfs may not follow the time dependence proposed by \citet{Wood}. For example, \citet{Silvestri} found that the lower mass M dwarfs (spectral types M3 and later) are more likely to be active at old ages than higher mass M dwarfs (M0-M3). M stars in clusters with spectral types M3 and later seem to
remain active for more than 4 Gyr \citep{Silvestri}. \cite{Gizisetal} found that late type M dwarfs (later than M6.5) that are strong H$\alpha$ emitters are likely to be old ($>$1 Gyr). Then, for the analysis of our results we will consider that the time dependent mass loss rate applies to earlier M dwarfs (earlier than M5) and the constant mass loss rate fits better the behavior of the later spectral type M dwarfs (later than M5). For these late type stars the percentage of active stars goes from $\sim$40\% for M5 stars to $\sim$60\% for M7 stars \citep{Silvestri, Gizisetal}.

For those early type M dwarfs, the atmosphere of a planet in their HZ may be safe for stellar mass loss rates 
$\leq 5 \times 10^3{\dot M}_\odot = 5 \times 10^{-11}$ M$_{\odot}$ yr$^{-1}$ for $\alpha=0.01$ (Figs. 2 and 4). For $\alpha=0.3$, only planetary atmospheres with  more than 5 atm may be able to survive for 1 Gyr or more if the parent star of the planet has mass losses $< 1 \times 10^3{\dot M}_\odot = 1 \times 10^{-11}$ M$_{\odot}$ yr$^{-1}$ (Figs. 3 and 5). Because most of early type M dwarfs are not active \citep{Silvestri, Gizisetal} we would expect that their mass loss rates would be less than the observed $\sim 10^{-11}$ M$_{\odot}$ yr$^{-1}$ for the most active M dwarfs (YZ CMi and AD Leo). We therefore expect that most planets in the habitable zone of early type M dwarfs may keep their atmospheres for 1 Gyr or more.

For M dwarfs that do not have a mass loss rate that
decreases as a function of time our time-independent wind model predicts the worst scenario for planets in the habitable zone of these stars.
If $\alpha=0.01$ a planet must have an atmosphere of 5 atm or more in order in order to survive mass loss rates of $10^{-11}$ M$_{\odot}$ yr$^{-1}$ (Figs. 2 and 4). For $\alpha=0.3$ (regardless of the planetary mass/atmospheric mass values) a planet in the HZ of an active late type M dwarf will loose its atmosphere in 1 Gyr or less (Figs. 3 and 5).

\section{Conclusions}

Because the habitable zone of M dwarfs is very close to the star ($\leq$2 AU) planets in that HZ may be tidally locked and therefore slow rotators. As a consequence these planets may have weak magnetic fields \citep{Griessmeieretal}. In this work we have considered the extreme case of planets with no magnetic field.

Our model gives the planetary atmosphere mass loss rate as a function of the stellar wind and planetary properties (mass, atmospheric pressure and orbital distance) and an entrainment efficiency coefficient $\alpha$. The model is applied to explore two different cases: a time-independent stellar mass loss and a stellar mass loss rate that decreases with time.
We find that our time-independet stellar wind model predicts similar results to the ones found by using detailed models of the atmospheric erosion of Venus. Therefore, our model is valid to test if a planet is able to keep its atmosphere for a certain period of time. In other words, if one planet (under the assumption that the planet has no magnetic field) doesn't pass our test, it is not necessary to analyze the atmospheric erosion with more detailed models.

Based on observations of M dwarfs we derive that our time dependent model may better describe early type M dwarfs (earlier than M5) and the case for constant stellar mass loss more appropriate for the interaction of M dwarfs with spectral types later than M5 with the atmospheres of planets in their habitable zones. Our results indicate that early type M dwarfs may are more promising for habitable planets searches because there are several planetary mass/atmospheric mass combinations for which a planet without magnetic field in the habitable zone may preserve its atmosphere for at least 1 Gyr. Planets in the habitable zone of active late type stars (later than M5) are less likely to keep their atmospheres for 1 Gyr or more.

It is clear that more observations are needed to narrow down the mass
loss rate values for M dwarfs. Particularly, such mass loss rate
determinations will have to be made for M dwarfs in which the presence
of planets is indeed detected, so as to be able to determine (from our
model) whether or not the planets are candidates for the presence of life.

We end the discussion by noting that our model for
the erosion of planetary atmospheres is restricted to the case
of planets with no magnetic fields. Clearly, the presence of
a planetary magnetosphere would protect the atmosphere, and may allow
the planet to preserve its atmosphere for longer timescales in the
presence of stellar winds with higher mass loss rates. The study
of the survival of the atmosphere of a magnetized planet under
the interaction with an M dwarf wind is left for a future study.

\section{Acknowledgments}
JZ thanks the support from CONACyT fellowship number 215794/207524.
This work was supported by the CONACyT projects 51715, 61547 and 79744, and PAPIIT project No. IN119709.

\end{document}